# SeisDiff-deno: A Diffusion-Based Denoising Framework for Tube Wave Attenuation in VSP Data

Donglin Zhu[1], Peiyao Li[1], and Ge Jin[1]

1. Department of Geophysics, Colorado School of Mines, Colorado, USA. E-mail: dzhu@mines.edu, lipeiyao@mines.edu, gjin@mines.edu

ABSTRACT

Tube waves present a significant challenge in vertical seismic profiling data, often obscuring critical seismic signals from seismic acquisition. In this study, we introduce the Seismic Diffusion Model for Denoising, a fast diffusion model specifically designed to remove the noise from seismic shotgather effectively. Our approach balances computational efficiency with high-quality image denoising, ensuring that the method is practical and robust for real-world applications. We validate the effectiveness of the proposed method through rigorous testing on both synthetic and field data, demonstrating its capability to preserve essential seismic signals while eliminating unwanted coherent noise. The results suggest that the proposed method enhances data quality and supports continuous production during seismic acquisition, paving the way for improved subsurface monitoring and analysis.

INTRODUCTION

Tube wave is a specific type of guided wave that propagates along a spherical fluid-solid interface, such as a fluid-filled borehole or pipeline. It can be observed in well downhole measurements like vertical seismic profiling (VSP) for obtaining high-resolution images (Cheng and Toksöz, 1981) and distributed acoustic sensing (DAS) records during completion of unconventional wells (Schumann and Jin, 2020). Depending on the applications, tube wave can be considered either as signal or as noise. As a signal, it can be used to characterize near-wellbore fracture properties (Hardin and Toksoz, 1985; Li et al., 1994; Bakku et al., 2013; Liang et al., 2017; Hunziker et al., 2020; Schumann and Jin, 2020; Zhang et al., 2021). However, in other cases, tube waves have been considered a problematic noise source for those trying to measure material properties in the medium surrounding the borehole (Pham et al., 1993; Herman et al., 2000; Daley et al., 2003). For example, without suppression or removal of the tube waves, reflections in VSP data can only be observed at large source-receiver distances before the tube waves arrive (Coates, 1998; Hunziker et al., 2020).

DAS VSP has seen rapid development in recent years as an efficient and cost-effective alternative to traditional geophone-based VSP systems. DAS technology relies on fiber optic cables, often installed along production tubing or casing, to capture seismic signals at fine spatial intervals, offering significant advantages in data acquisition and well monitoring. One major benefit is the ability to use pre-existing fiber optic infrastructure, eliminating the need for additional downhole equipment and minimizing the logistical challenges associated with deploying traditional geophones. Reducing rig time and operational complexity can result in substantial cost savings, making DAS an attractive option for many operators(Mateeva et al., 2013; Jousset et al., 2018).

DAS installations can be categorized based on their placement within the wellbore, primarily as tubing DAS and casing DAS configurations. In tubing DAS installations, the fiber optic cable is attached or clamped along the production tubing. This configuration allows for close monitoring of the wellbore environment, making it particularly effective for detecting seismic signals and other downhole activities. Tubing DAS provides higher sensitivity to wellbore events due to its proximity to fluid flow and the production tubing. However, it also presents challenges, such as increased noise from production activities, especially when fluid movement generates strong tube waves during injection or production operations (Daley et al., 2013). In contrast, casing DAS installations embed the fiber optic cable along the casing. While this setup is less sensitive to wellbore dynamics, it experiences lower noise levels and has longer-term durability, making it more suitable for continuous reservoir monitoring.

Despite these technical advantages, fiber installations—whether along tubing or casing—pose specific challenges, particularly regarding data acquisition during active production or injection operations. Unlike traditional geophone-based VSP, often conducted during shut-in periods to minimize fluid-related noise, DAS VSP operates in real time, even during active production or injection. This introduces stronger tube waves, created by fluid movement along the wellbore, leading to higher noise levels in the recorded data. The intensity of tube waves during production or injection operations is much stronger than the quieter conditions in shut-in operations typically used for geophone-based measurements. These waves can mask the seismic signal of interest, requiring sophisticated processing and noise suppression techniques to accurately interpret the data (Daley et al., 2013).

Suppression of tube waves can be implemented during data processing (Hardage, 1981; Herman et al., 2000; Daley et al., 2003; Greenwood et al., 2019). Tube waves appear as linear events with strong amplitude, which can override desired signals (Houston, 1992). Frequency–wavenumber (FK) filters are often used to suppress tube waves before analyzing body waves (Afanasiev et al., 2014; Nakata et al., 2022). However, the process of filtering out certain frequency-wavenumber components might inadvertently remove some relevant seismic signals, especially if the slowness of tube waves overlaps significantly with the desired seismic reflections. Wavelet-based methods like median filters can also be used to suppress tube waves (Houston, 1992). However, the performance of the median filter is sensitive to the size of the window used for filtering. An incorrectly chosen window size can either inadequately suppress tube waves or over-smooth the data (Houston, 2005).

The limitations of these filters and the requirements for high-quality seismic images motivate researchers to explore advanced methods. Convolutional neural networks (CNN) have been applied to suppress different kinds of noise for VSP in both geophone and DAS records (Cheng et al., 2023; Guo et al., 2023; Yang et al., 2023a, 2023b, 2023c; Dong et al., 2022; Zhao et al., 2022). The generative models, which show better robustness to variations in noise and sophisticated noise modeling, have recently drawn much attention. Generative adversarial networks are introduced to remove noise and restore weak seismic events simultaneously (Wu et al., 2023). Diffusion models (Sohl-Dickstein et al., 2015; Ho et al., 2020) are applied to suppress coupling noise from DAS-VSP data (Zhu et al., 2023). However, there are few studies on effectively suppressing tube waves through deep learning methods.

To enhance image quality and preserve signals on VSP images, we propose a diffusion model specifically designed for tube wave suppression. The training process involves generating volumes of VSP shot gathers, followed by an efficient workflow to simulate tube waves and incorporate them into the field shot gathers. The success of the synthetic tests prompted a field test using the trained, fast, and improved conditional diffusion model. The results from both synthetic and field tests demonstrate the effectiveness of the proposed approach. Additionally, we emphasize the model's strong generalization ability by successfully applying the field-trained model to synthetic data containing simulated tube waves and extracted field noise.

## METHODOLOGY

This section illustrates the workflow of the proposed diffusion model for tube wave suppression. We show the tube wave simulation procedure for VSP data. Then, we define the generating method of training dataset for two scenarios: synthetic VSP data with simulated tube wave and field VSP records with simulated tube wave. A fast improved conditional diffusion model is introduced for the tube wave suppression task. Finally, we describe the training and inference strategies.

**Tube wave simulation**

Simulating tube waves as linear noise in VSP data is a nuanced process that involves careful manipulation of seismic noise to mimic the characteristics of tube waves. Tube waves, being guided waves that travel along the borehole in fluid-filled environments, have distinct properties such as lower frequency ranges, specific velocities, and unique attenuation patterns. The simulation begins with a detailed analysis of the VSP data to isolate and characterize the linear noise component. The tube wave can be observed from VSP data as linear noise with certain slope across the entire shotgathers. Linear noise, known for its predictability and consistency, is an ideal simulation base due to its waveform stability.

The key to the simulation is aligning the characteristics of the linear noise with those of the tube waves. This process involves estimating the tube wave velocity and calculating the dip based on the velocity and channel spacing. A flexible range of velocity may be applied to obtain higher generalization than a fixed velocity. Additionally, scaling the amplitude of the linear noise to mirror the amplitude range of tube waves is crucial for realism. The tube wave can be simulated by the following:

$$L(x, y) = Random\ (x), \tag{1}$$

$$N_{tube} = A\ L(x, y)\ R(\theta), \tag{2}$$

$$R = \begin{bmatrix} \cos\theta & -\sin\theta & 0 \\ \sin\theta & \cos\theta & 0 \\ 0 & 0 & 1 \end{bmatrix}, \tag{3}$$

$$N_{total} = s_1\ N_{tube-up} + s_2\ N_{tube-down}, \tag{4}$$

where $N$ represents to the noise, $R$ is the rotation matrix, $\theta$ is the dip calculated by tube wave velocity and geometry of the survey, and $A$ denotes the amplitude scaler.

Further refinement includes amplitude changes, where the linear noise is modified to adopt a mask on simulated tube waves. Once the linear noise closely resembles field tube waves, it is reintegrated into the VSP data. This step is delicate, as it involves superimposing or replacing parts of the original data with the modified noise, ensuring a seamless blend that maintains the integrity of the original data.

**Field VSP shot gathers training data**

In VSP surveys, geophones or fiber optic cables are typically installed in wells for a long time, especially for DAS-VSP, which might be installed permanently. This setup allows for continuous monitoring. Consequently, the field dataset recorded during the down periods when the well is not in active production is devoid of tube waves, though the dataset may still contain other types of noise.

To create a training dataset suitable for our denoising model, we start with relatively clean (tube wave free) field records. Following our established procedure, we introduce simulated tube waves to this clean data. This approach generates the desired noisy data while retaining the original tube wave-free recordings as reference clean data (Figure 1). This method ensures we have a comprehensive set of noisy-clean data pairs for effectively training the model.

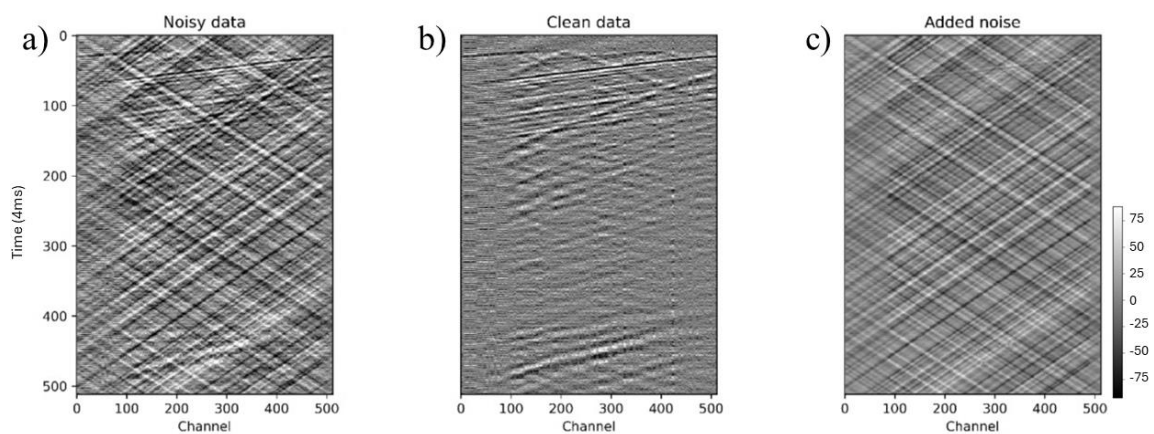

Figure 1. Training data example, a) noisy data, b) clean data, and c) noise added to clean data

**Seismic Diffusion Model for Denoising**

To suppress seismic noise, we adopted and modified the Seismic Diffusion Model for Denoising (SeisDiff-deno) approach based on the description by (Song et al., 2020), (Nichol and Dhariwal, 2021), and (Jiang et al., 2024). Our method utilizes a diffusion model to improve denoising performance while significantly reducing computational costs for both training and inferring. The workflow is shown in Figure 2.

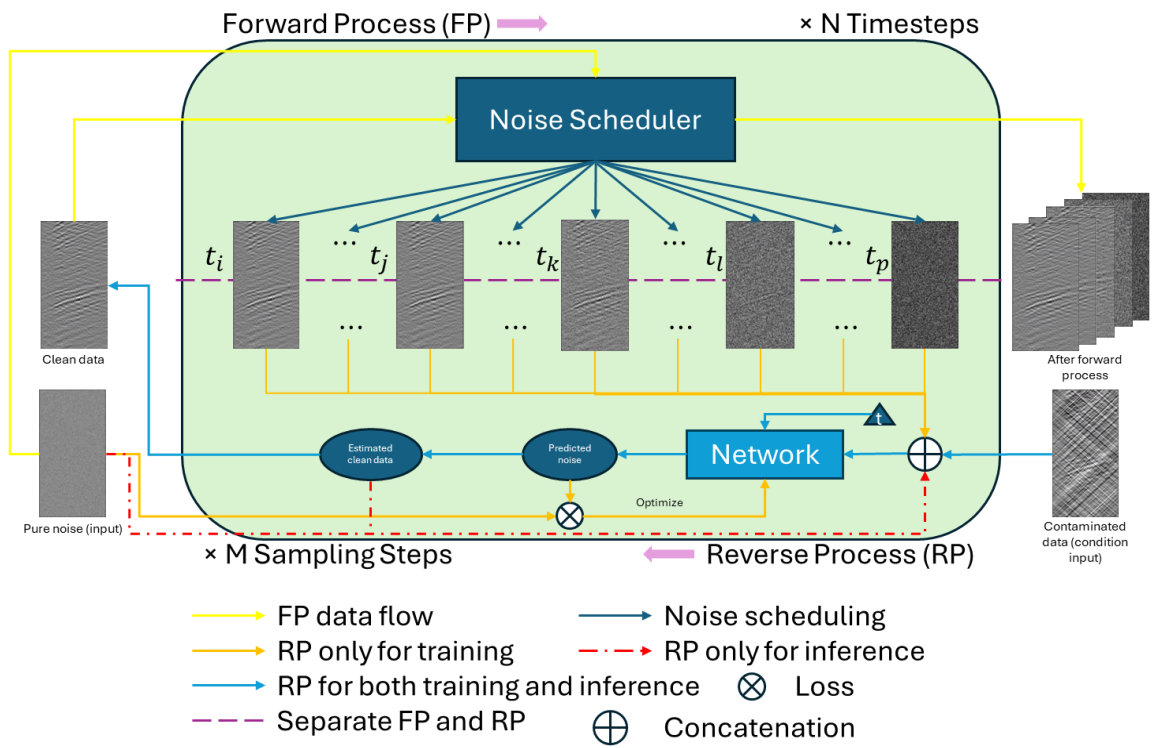

Figure 2. SeisDiff-deno workflow

*Conditional Denoising Diffusion Probabilistic Model (DDPM)*

The conditional DDPM consists of two phases: a forward process and a reverse process, with an additional input to control the generated image. The forward diffusion process progressively adds Gaussian noise to the original seismic signal $x_0$, transforming it into pure Gaussian noise $x_T$ over T time steps. The reverse process, parameterized by a neural network, aims to reconstruct the original signal by removing the noise step-by-step.

The forward process is defined as:

$$q(x_t|x_{t-1}) = N(x_t; \sqrt{1-\beta_t}x_{t-1}, \beta_t \mathbf{I}), \tag{5}$$

where variance $\beta_t \in (0,1)$, $N$ denotes normal distribution, $\mathbf{I}$ represents the unit matrix. The noising process defined in Equation 1 allows us to sample an arbitrary step of the noised latents directly conditioned on the input $x_0$ (Ho et al., 2020; Nichol and Dhariwal, 2021). With $\alpha_t = 1 - \beta_t$, $\bar{\alpha}_t = \prod_{i=1}^{t} \alpha_i$, and Gaussian noise $\epsilon$, the equation 1 can be rewritten as:

$$q(x_t|x_0) = N(x_t; \sqrt{\bar{\alpha}_t}x_0, (1-\bar{\alpha}_t)\mathbf{I}), \tag{6}$$

$$x_t = \sqrt{\bar{\alpha}_t}x_0 + \sqrt{1-\bar{\alpha}_t}\epsilon, \tag{7}$$

The reverse process is defined as:

$$p_\theta(x_{t-1}|x_t, c) = N(x_{t-1}; \mu_\theta(x_t, c, t), \Sigma_\theta(x_t, c, t)\mathbf{I}), \tag{8}$$

$$\mu_\theta(x_t, c, t) = \frac{1}{\sqrt{\bar{\alpha}_t}}\left(x_t - \frac{\beta_t}{\sqrt{1-\bar{\alpha}_t}}\epsilon_\theta(x_t, c, t)\right), \tag{9}$$

$$\Sigma_\theta(x_t, c, t) = \sigma_t^2 = \frac{1-\bar{\alpha}_{t-1}}{1-\bar{\alpha}_t}\beta_t, \tag{10}$$

where $\theta$ represents the parameters learned by the neural network, $\epsilon_\theta$ is the output of the network, $c$ denotes the condition input. The neural network used in the reverse process to simulate the distribution $p_\theta(x_{t-1}|x_t, c)$ is the U-Net (Ronneberger et al., 2015) with ResNet blocks (He et al., 2015) and self-attention (Vaswani et al., 2017). The network (Att-ResUnet) could predict the added noise $\epsilon_\theta$ by optimizing loss function (Ho et al., 2020):

$$L_t = E_{t,x_0,c,\theta}\left[\left\|\epsilon - \epsilon_\theta\left(\sqrt{\bar{\alpha}_t}x_0 + \sqrt{1-\bar{\alpha}_t}\epsilon, t, c\right)\right\|^2\right], \tag{11}$$

Then, the denoised image can be calculated by:

$$x_{0_{from\epsilon_\theta}} = \sqrt{\frac{1}{\bar{\alpha}_t}}x_t - \sqrt{\frac{1}{\bar{\alpha}_t}-1}\epsilon_\theta(x_t, c, t). \tag{12}$$

*Improving the Network Architecture*

The network architecture (Figure 3) is enhanced with ResNet blocks and self-attention mechanisms for efficient and high-quality image denoising. It starts with a 2D input layer, followed by a timestep embedding module that uses sinusoidal positional encoding. The downsampling path includes multiple levels, each with ResNet blocks and optional attention blocks, and may include a Downsample layer to halve the spatial dimensions. The middle block operates at the lowest resolution and contains ResNet and attention blocks. The upsampling path mirrors the downsampling path, with ResNet blocks, optional attention blocks, and skip connections that concatenate feature maps from corresponding downsampling levels, along with optional Upsample layers to double the spatial dimensions. The final output layer consists of normalization followed by a convolutional layer that generates the output image. Key components include GroupNorm, Swish activation, dropout within ResNet blocks, scaled dot-product attention, and temporal embedding to condition the model on diffusion timesteps. The model is configured with parameters such as base and output channels, channel multipliers, number of ResNet blocks, attention resolutions, dropout rate, input channels, image resolution, and optional convolutional resampling. This architecture effectively combines U-Net, ResNet, and attention mechanisms to adapt to different timesteps and produce high-quality denoised images.

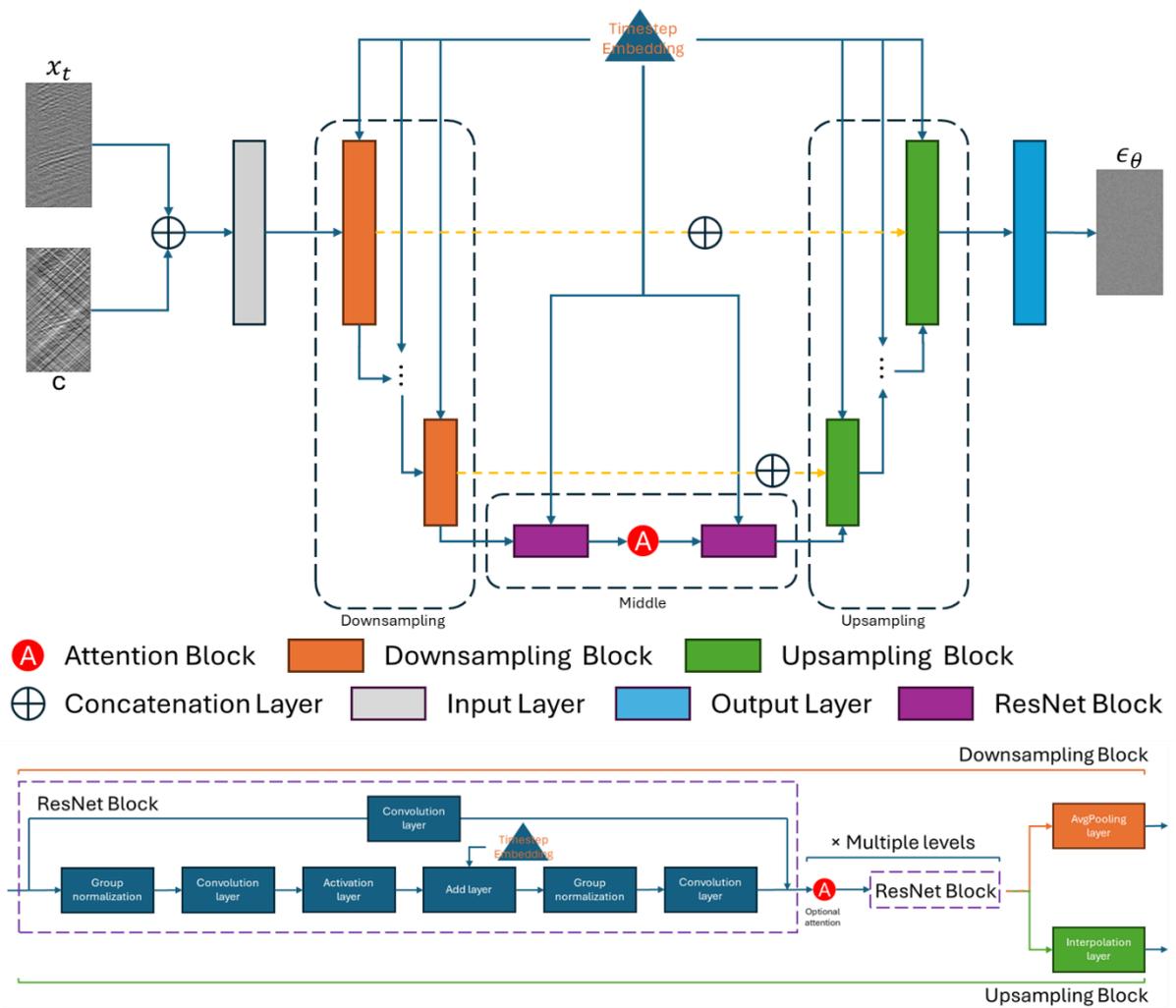

Figure 3. Network architecture in the reverse process

*Denoising Diffusion Implicit Model (DDIM) Sampler*

One problem with the DDPM process is the speed of generating an image after training is done. The DDIM (Song et al., 2020) introduces a way to speed up image generation with little image quality tradeoff. It does so by redefining the diffusion process as a non-Markovian process, which allows for skipping steps in the denoising process, not requiring all past states to be visited before the current state. DDIMs can be applied after training a model, so DDPM models can easily be converted into a DDIM without retraining a new model. The reverse diffusion process for a single step is redefined:

$$x_{t-1} = \sqrt{\bar{\alpha}_{t-1}} \left( \frac{x_t - \sqrt{1-\bar{\alpha}_t}\epsilon_\theta}{\sqrt{\bar{\alpha}_t}} \right) + \sqrt{1 - \bar{\alpha}_t - \sigma_t^2} \cdot \epsilon_\theta + \sigma_t \epsilon, \tag{13}$$

$$\sigma_t(\eta) = \eta \sqrt{\frac{1-\bar{\alpha}_{t-1}}{1-\bar{\alpha}_t}} \sqrt{\frac{1-\bar{\alpha}_t}{\bar{\alpha}_{t-1}}} = \eta \sqrt{\frac{1-\bar{\alpha}_{t-1}}{1-\bar{\alpha}_t} \beta_t}, \tag{14}$$

The diffusion model is a DDIM when $\eta = 0$ as there is no noise and an original DDPM when $\eta = 1$. Any $\eta$ between 0 and 1 is an interpolation between a DDIM and DDPM. We set $\eta = 0$ to obtain deterministic results, ensuring that the output remains consistent under identical conditions, which is crucial for seismic denoising tasks.

*Improving the Noise Schedule*

The noise schedule plays a critical role in diffusion models. In particular, the later stages of the forward process become overly noisy, which adds little to the final samples' quality when using a linear scheduler (Nichol and Dhariwal, 2021). Essentially, noise is being added too fast. Instead, the cosine scheduler adds noise slower to retain image information for later time steps. A cosine schedule is designed to have a linear drop-off of $\bar{\alpha}_t$ in the middle of the process, while changing very little near the extremes of t = 0 and t = T to prevent abrupt changes in noise level (Figure 4). The noise schedule in terms of $\bar{\alpha}_t$ can be constructed as:

$$\bar{\alpha}_t = \frac{f(t)}{f(0)}, \tag{15}$$

$$f(t) = \cos\left( \frac{\frac{t}{T}+s}{1+s} \cdot \frac{\pi}{2} \right)^2, \tag{16}$$

where $s$ is a small offset to prevent the value too small near $t = 0$.

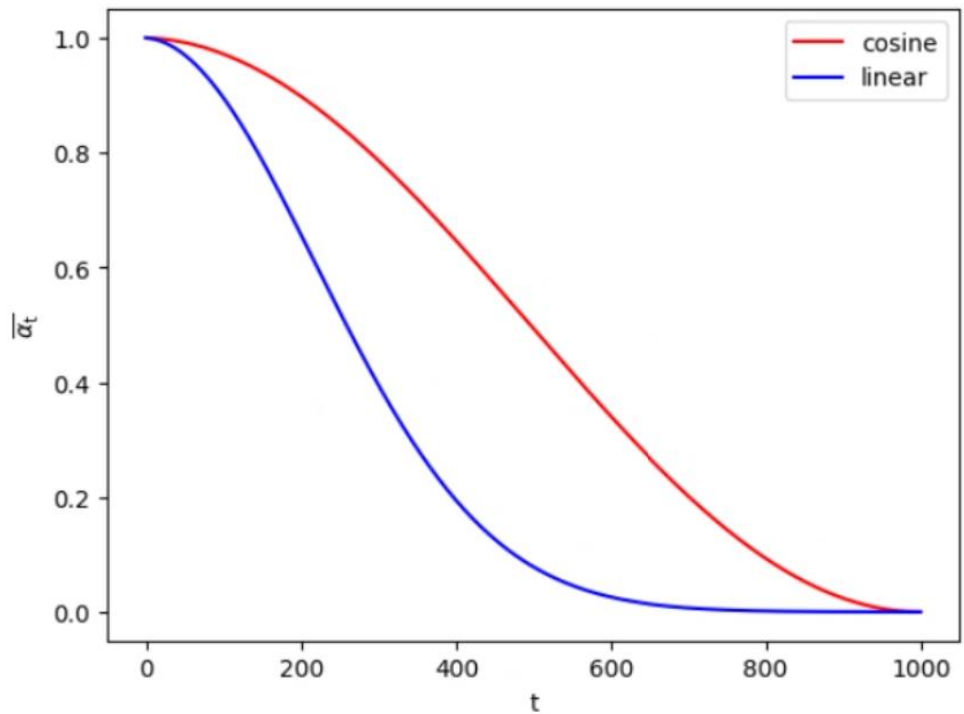

Figure 4. Impact of the number of time steps $t$ on the noise schedule

DDPM demonstrates that the total number of time steps $T = 1000$ is the best choice in most cases for high-quality image generation. Too large or too small time steps reduce the performance of the DDPM (Ho et al., 2020). However, using large time steps like 1,000 usually requires several days for training and minutes to hours for inferring (sampling), even for a small size of images. DDIM completes the reverse process using ordinary differential equations (ODE), significantly reducing time to much smaller time steps (e.g., we use 50 in this study) without impacting generation quality.

**Training and inference**

The model was trained on a computing cluster equipped with NVIDIA A100 GPUs. The training batch size was set to 16, and the AdamW optimizer with a learning rate of

$2 \times 10^{-4}$ was used. The training process involved 20,000 iterations for SeisDiff-deno. The exponential moving average (EMA) is applied to obtain a stable and robust model. Detailed training parameters are listed in Table 1. The inference is conducted on NVIDIA RTX A4000 with 16GB memory.

Tabel 1. Parameters for the SeisDiff-deno training

| Name | Parameters |
| --- | --- |
| Input Channel | 2 |
| Output Channel | 1 |
| Number of ResBlock | 2 |
| Attention Resolution | 16 |
| Base Filter Number | 128 |
| Multiple of Filter Num | [1, 1, 2, 2, 4, 4] |
| EMA Rate | 0.999 |
| Diffusion Timestep | 1000 |
| Sampling Step | 50 |

APPLICATION

We introduce synthetic testing for quantitative evaluation. We use the signal-to-noise ratio (SNR) to evaluate the denoising results for testing quantitively, the cross-correlation coefficient (CCE) to measure the signal leakage, and the structural similarity index measure (SSIM) to evaluate the similarity between the denoised image and the ground truth image (Wang et al., 2004) if the ground truth is known. The SNR is defined as:

$$SNR = 10 log_{10} \left(\frac{P_{signal}}{P_{noise}}\right), \tag{17}$$

where $P_{signal}$ is the power of the signal, and $P_{noise}$ is the power of the noise. The cross-correlation coefficient can be defined by:

$$CCE(x, \bar{x}) = \frac{\sum_{i=1}^{N}(x_i - \bar{x})(y_i - \bar{y})}{\sqrt{\sum_{i=1}^{N}(x_i - \bar{x})^2 \sum_{i=1}^{N}(y_i - \bar{y})^2}}, \tag{18}$$

where $\bar{x}$ and $\bar{y}$ are the mean value of clean data and removed noise, respectively. The SSIM can be defined:

$$SSIM(x, \bar{x}) = \frac{(2\mu_x \mu_{\bar{x}} + c_1)(2\sigma_{x\bar{x}} + c_2)}{(\mu_x^2 + \mu_{\bar{x}}^2 + c_1)(\sigma_x^2 + \sigma_{\bar{x}}^2 + c_2)}, \tag{19}$$

where $\mu_x$ and $\mu_{\bar{x}}$ are the mean values of clean data and denoised data, respectively. $\sigma_x^2$ and $\sigma_{\bar{x}}^2$ are the mean values of clean data and denoised data, respectively. $\sigma_{x\bar{x}}$ represents the covariance between clean data and denoised data. $c_1$ and $c_2$ are small constants to stabilize the division when the denominator is close to 0.

**Testing**

To better illustrate the advantages of the proposed method in handling wavefields blended with tube waves, we introduce a synthetic testing scenario. We use elastic VSP data from SEAM Phase I (Fehler and Larner, 2008) as the clean dataset.

Linear noise is generated and added according to the proposed workflow, ensuring that the linear noise shares the same dip as part of the wavefield for this specific evaluation. This setup makes it challenging for the FK filter to distinguish between the signal and noise, inevitably leading to the removal of noise and portions of the signal. However, the proposed SeisDiff-deno method can easily reconstruct the signal hidden within the blended noise, demonstrating its superior capability in preserving the integrity of the wavefield (Figure 5). The quantitative comparison is listed in Table 2. For better visualization, Figure 6 and Figure 7 show the zoomed-in part from Figure 5.

Table 2. Quantitively comparison between FK filter and SeisDiff-deno on synthetic test

| Method | SSIM | SNR (dB) |
|---|---|---|
| FK Filter | 0.49 | 12.97 |
| SeisDiff-deno | 0.92 ↑ | 18.58 ↑ |

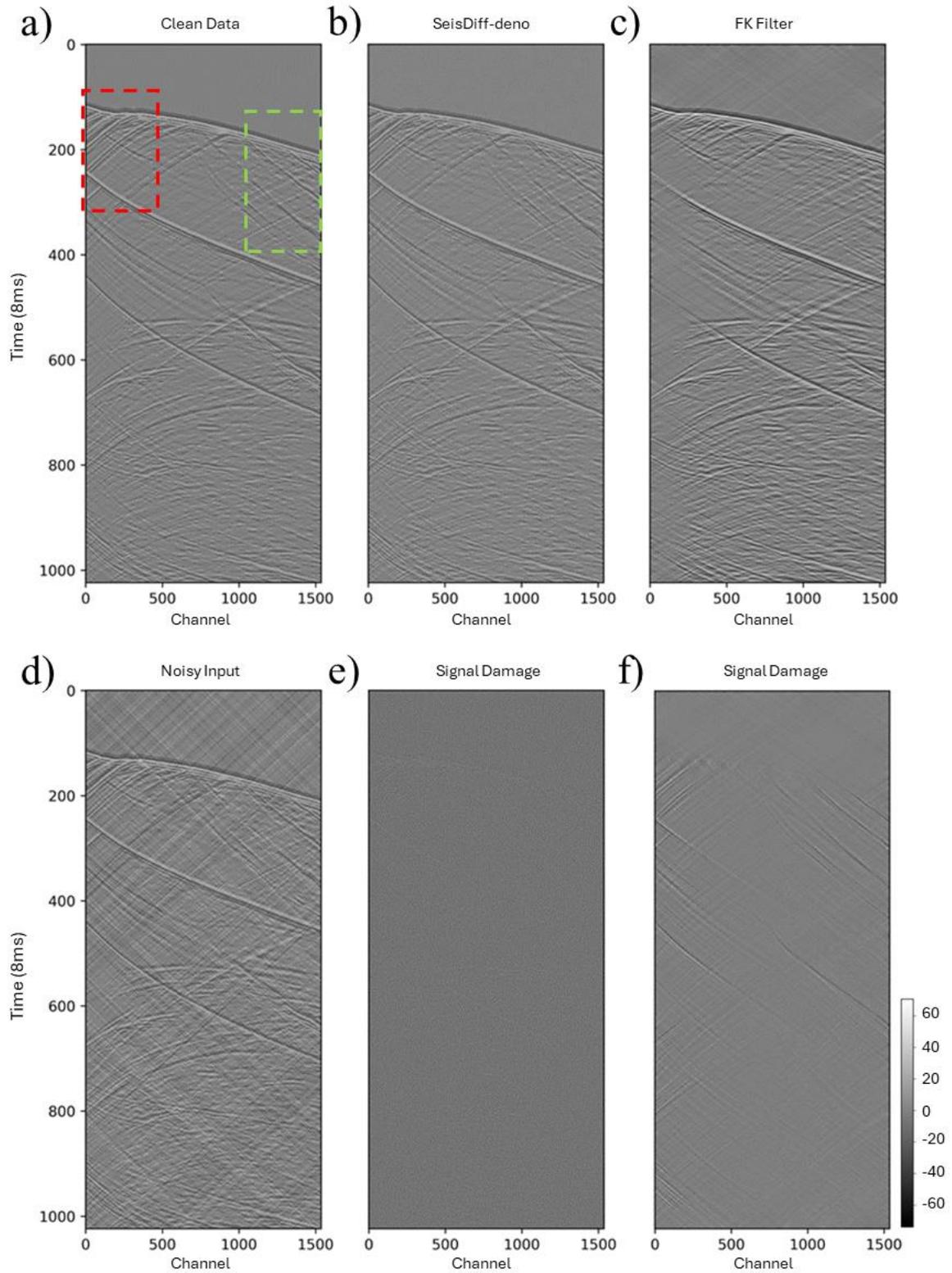

Figure 5. Synthetic test on SEAM model, a) clean data, b) denoising result from SeismDiff-deno, c) denoising result from FK filter, d) noisy input, e) residual between a) and b), f) residual between a) and c).

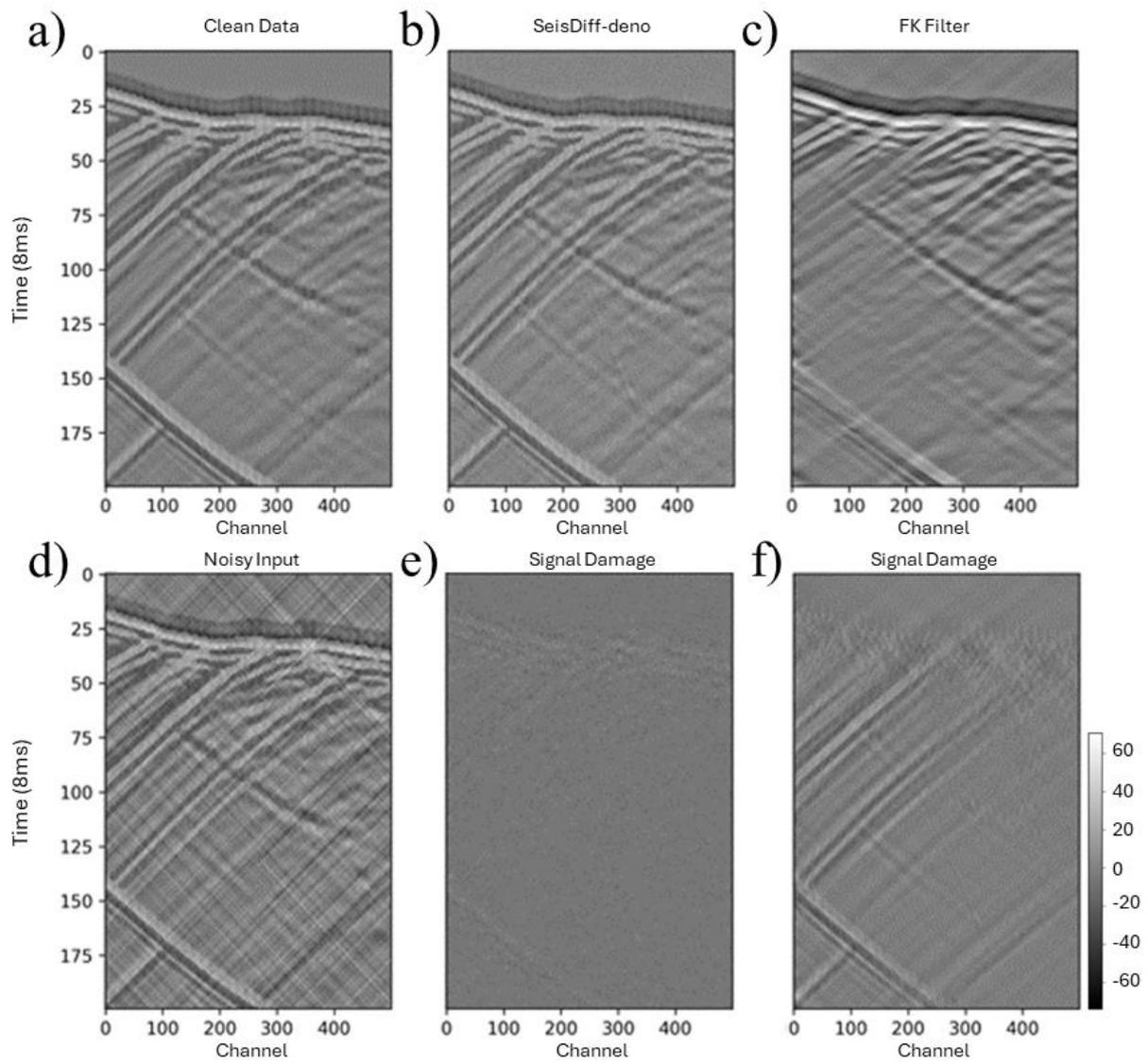

Figure 6. Zoomed in part of the red box on Figure 5. a) clean data, b) denoising result from SeismDiff-deno, c) denoising result from FK filter, d) noisy input, e) residual between a) and b), f) residual between a) and c).

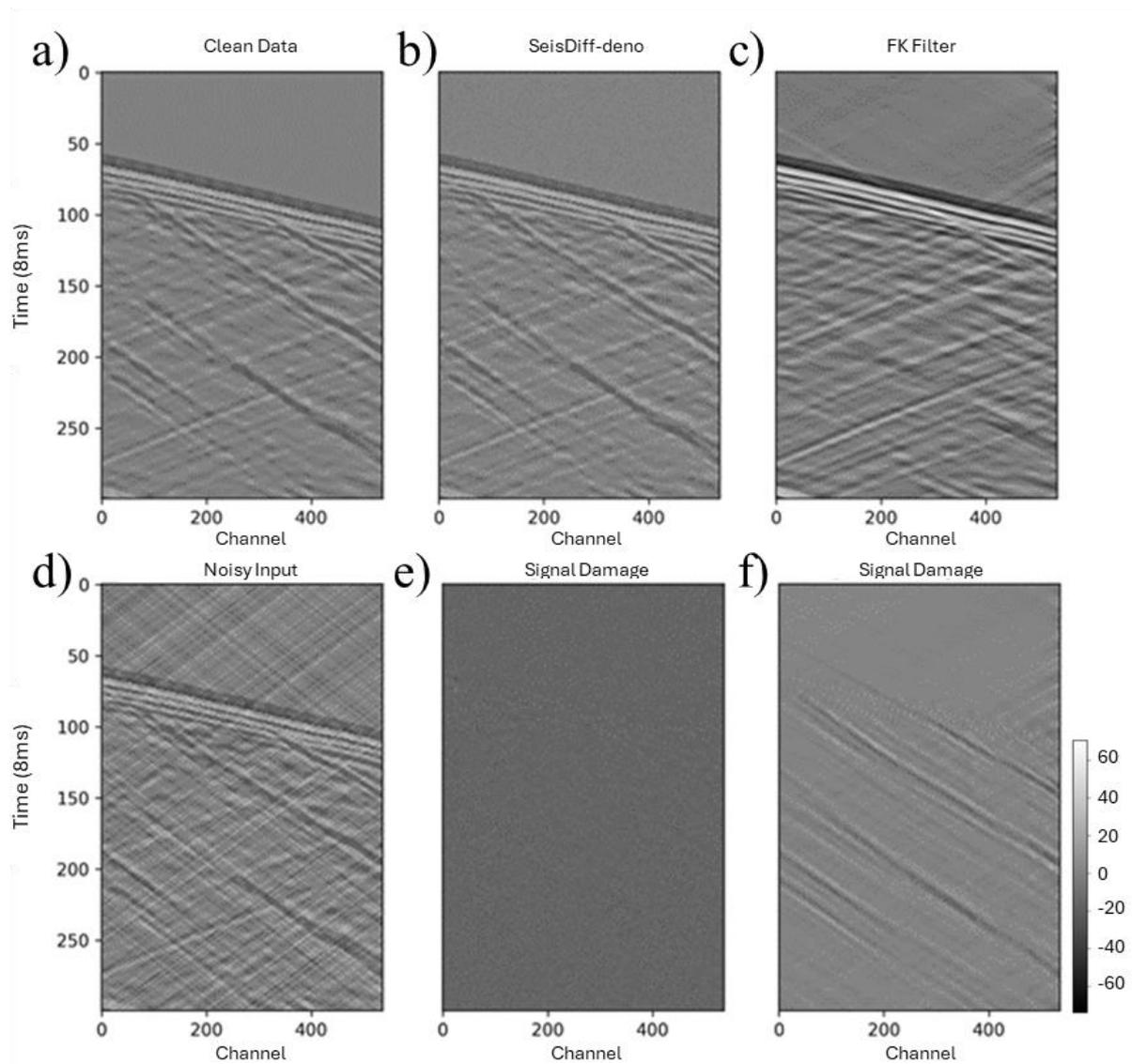

Figure 7. Zoomed in part of the green box on Figure 5. a) clean data, b) denoising result from SeismDiff-deno, c) denoising result from FK filter, d) noisy input, e) residual between a) and b), f) residual between a) and c).

## DISCUSSION

*Generalization of the proposed method*

In our study, we apply the trained model on field VSP data to remove noise from a synthetic dataset derived from the SEAM model characterized by a different geological setting.

This approach underscores the proposed method's robustness, as it effectively attenuated the noise without being specifically trained on clean data or requiring prior knowledge of the velocity model underlying the synthetic data. The success of this test demonstrates the proposed SeisDiff-deno' 's adaptability and generalization capability, indicating that it can perform well across diverse geological conditions, even when there is a significant disparity between the training and application datasets (Figure 8 and Figure 9). Figure 8 presents SEAM VSP data with simulated tube waves, while Figure 9 shows SEAM VSP data with extracted field noise from another field DAS-VSP dataset. This field data is a 4D DAS-VSP dataset from a deepwater field in the Gulf of Mexico (GoM). The GoM VSP data were recorded by the multi-mode fiber optical cables located in two wells, both active injector wells, from 2015 to 2018 (Zwartjes et al., 2018). The noise is captured from the GoM dataset above the first arrival. A quantitative comparison of the results for Figure 9 is provided in Table 3. The correlation coefficient serves as a "smaller is better" metric for measuring signal leakage. Based on the results, we observe that the proposed method achieves a significantly lower correlation coefficient while providing a higher SNR compared to the FK filter.

This outcome illustrates that the specific choice of velocity model or geological setting is not critical for denoising. Instead, the proposed method effectively learns the data distribution inherent in the data and is controlled by the conditional input, enabling them to generalize across various geological scenarios. This robustness underscores the potential of our denoising approach to be applied broadly, regardless of the underlying geological complexities. By focusing on the noise characteristics, the proposed method demonstrates consistent performance, highlighting the efficacy of the methodology in diverse seismic data.

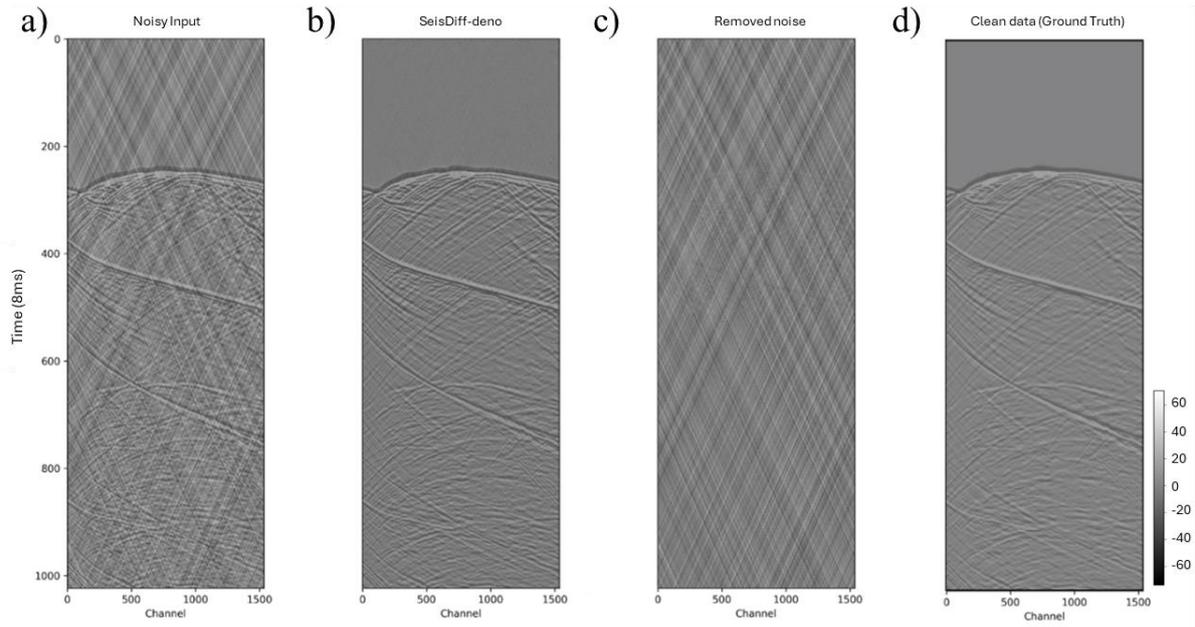

Figure 8. The denoising result on SEAM VSP record with simulated tube waves by using field data trained model. a) the noisy input (VSP with simulated tube waves), b) the denoising result from SeisDiff-deno, c) the removed noise, d) clean data (ground truth).

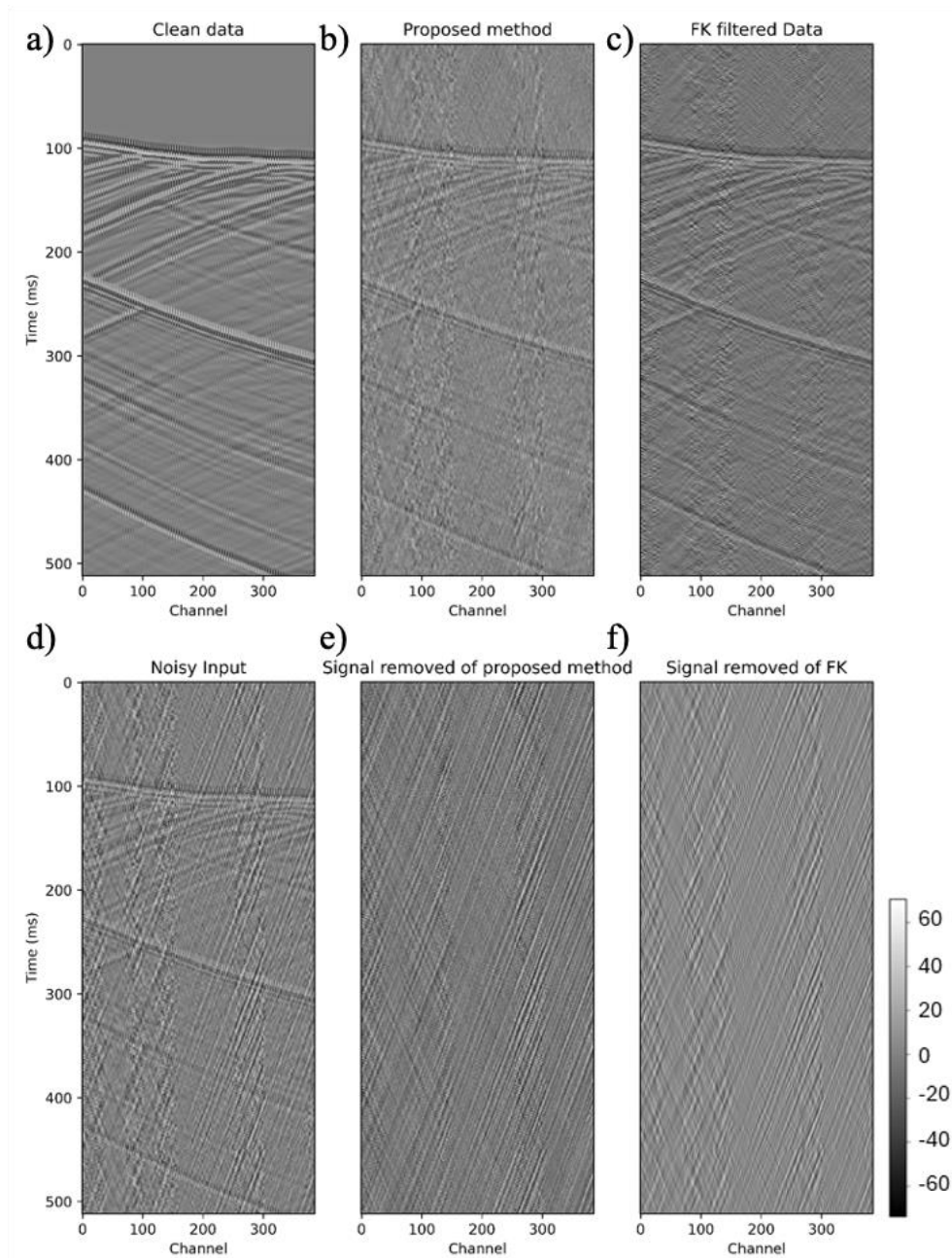

Figure 9. The denoising result on SEAM VSP records with extracted field noise by using field data trained model. a) clean data (ground truth)., b) denoising result from SeisDiff-deno, c) the removed noise, d) the noisy input (VSP with field noise), e) residual between b) and d), and f) residual between c) and d).

Table 3. Quantitively comparison between FK filter and SeisDiff-deno on generalization test

|  | Correlation coefficient | SNR |
|---|---|---|
| FK filter | 0.45 | 9.01 |
| SeisDiff-deno | 0.15 ↓ | 10.27 ↑ |

*Compare with the original DDPM*

In this study, the proposed SeisDiff-dono adopts the fast diffusion model for tube wave suppression, and we provide a comparative analysis of its performance against the original DDPM. The original DDPM, while highly effective in denoising tasks, operates through an extensive iterative process that reconstructs the signal by gradually reversing a noisy forward process. This method, though precise, is computationally intensive and often impractical for handling huge amounts of shot gathers. In contrast, the fast diffusion model powered by DDIM sampling introduces a deterministic approach to the reverse process, significantly reducing the number of required diffusion steps in inferring. This makes DDIM more computationally efficient while maintaining a comparable denoising quality, making it suitable for scenarios where speed is a critical factor.

This efficiency is particularly advantageous in seismic data processing, where large volumes of data must be handled swiftly and accurately. By adopting SeisDiff-deno, we ensure that our denoising process is effective in preserving the integrity of seismic signals and optimized for practical, real-world applications where computational resources and time are limited. This balance between denoising quality and processing speed makes SeisDiff-deno an ideal choice for seismic denoising tasks, providing a robust solution that meets both performance and efficiency requirements.

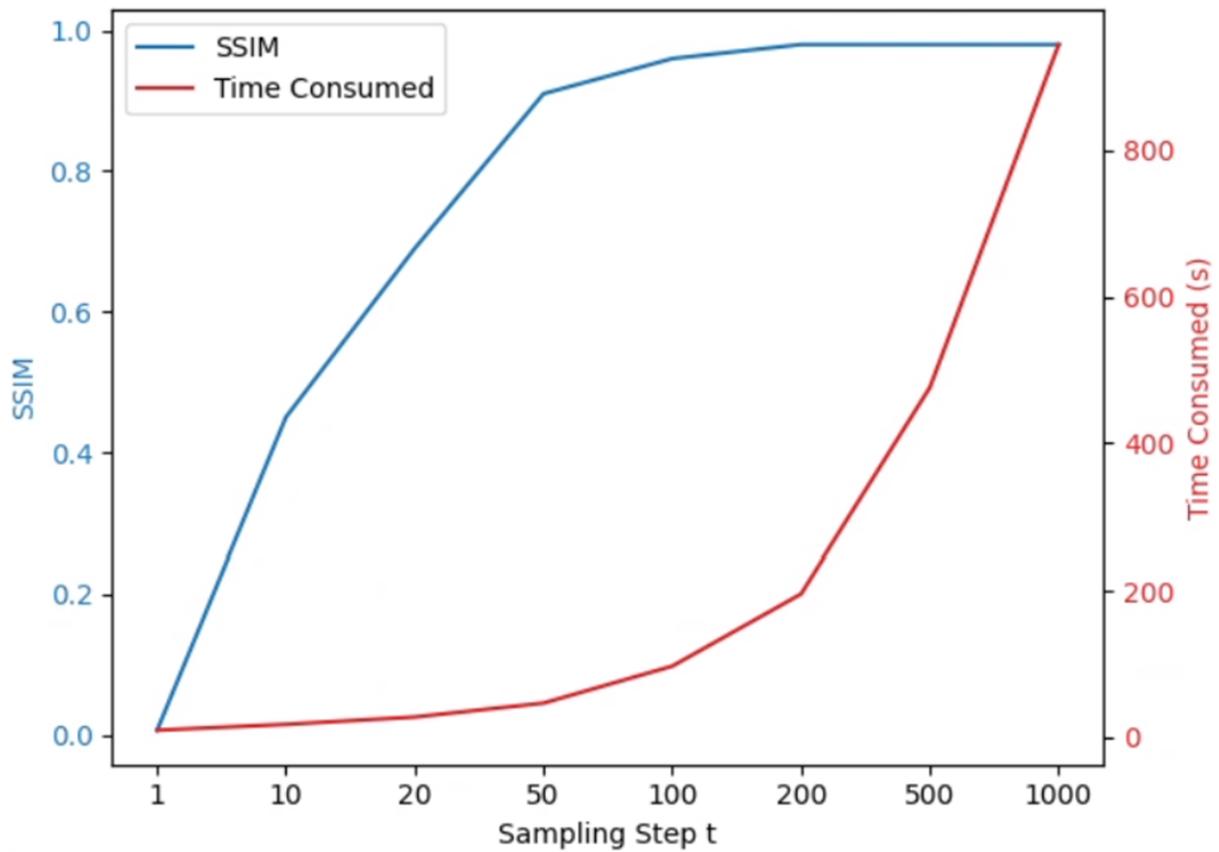

Figure 10. The SSIM and time consumption plot with different sampling steps. The SSIM and time consumption are calculated using a 1024 by 1536 synthetic shotgather.

CONCLUSION

In this study, we successfully employed SeisDiff-deno, a fast improved conditional diffusion model, to eliminate tube waves from VSP data. Testing on synthetic and field data demonstrated the model's robustness and generalizability across different geological settings. The proposed method has been demonstrated to effectively suppress the tube wave and recovers all other signals from the noisy input. Our results show that the proposed method can significantly enhance the quality of VSP recordings, making it feasible to acquire reliable seismic data even while wells are in production. This capability is particularly valuable for ongoing reservoir monitoring and management, as it allows for continuous data acquisition without disrupting production activities. Using SeisDiff-deno not only improves the efficiency of VSP data processing but also ensures high-quality seismic images, which are crucial for accurate subsurface characterization.